\documentclass{article} %
\usepackage{graphicx}
\usepackage{wrapfig}
\usepackage{enumitem}
\usepackage{amssymb}
\usepackage{color}
\newcommand{\EMS}[1]{\textcolor{red}{[\textbf{EMS: #1}]}}

\newcommand{\remove}[1]{}
\newcommand{\figSize}{0.25}

\newcommand{\reduceSpace}{-0pt}

\usepackage[linesnumbered,ruled,vlined,noresetcount]{algorithm2e}
\SetKwInput{KwVar}{Variables}
\SetKwInput{KwMacro}{macro}
\SetKwInput{KwFunc}{function}
\SetKwInput{KwEvent}{Event handler at server i}
\SetKwProg{Upon}{upon}{ do}{}
\SetKwProg{Doforever}{Do forever begin}{}{}

\usepackage{amsfonts,amsmath,url}
\usepackage{tabu}
\usepackage{diagbox}
\usepackage[square,numbers,comma,sort&compress]{natbib} 

\begin{document}
	%
	\title{Self-stabilization Overhead: an Experimental Case Study on Coded Atomic Storage\\\large{(preliminary version)}}
		
		\author{Chryssis Georgiou~\footnote{University of Cyprus, Cyprus \texttt{chryssis@cs.ucy.ac.cy}. $^\dagger$Chalmers University of Technology, Sweden \texttt{\{robg,lindhea\}@student.chalmers.se, elad@chalmers.se}} \and 
		 Robert Gustafsson~$^\dagger$ \and  Andreas Lindh\'{e}~$^\dagger$ \and Elad M.\ Schiller~$^\dagger$} 
	
	\maketitle  

\begin{abstract}
Shared memory emulation on distributed message-passing systems has attracted much attention over the past three decades. It can be used as a fault-tolerant and highly available distributed storage solution or as a low-level synchronization primitive. Examples of its uses can be found in cloud computing and cloud storage. Attiya, Bar-Noy, and Dolev were the first to propose a single-writer, multi-reader linearizable register emulation where the register is replicated to all servers. Many works followed; considering solutions for the multi-writer, multi-reader setting, as well as for supporting dynamic server participation. Recently, Cadambe et al. proposed the Coded Atomic Storage (CAS) algorithm, which uses erasure coding for achieving data redundancy with much lower communication cost than previous algorithmic solutions. 

Although CAS can tolerate server crashes, it was not designed to recover from unexpected, transient faults, without the need of external (human) intervention. In this respect, Dolev, Petig, and Schiller have recently developed a \emph{self-stabilizing} version of CAS, which we call CASSS. As one would expect, self-stabilization does not come as a free lunch; it introduces, mainly, communication overhead for detecting inconsistencies and stale information. So, one would wonder \emph{whether the overhead introduced by self-stabilization would nullify the gain of erasure coding}. 

To answer this question, we have implemented and experimentally evaluated the CASSS algorithm on PlanetLab; a planetary scale distributed infrastructure. The evaluation shows that our implementation of CASSS scales very well in terms of the number of servers, the number of concurrent clients, as well as the size of the replicated object. More importantly, it shows (a) to have only a \emph{constant} overhead compared to the traditional CAS algorithm (which we also implement) and (b) the recovery period (after the last occurrence of a transient fault) is as fast as a few client (read/write) operations. Our results suggest that CASSS does not significantly impact efficiency while dealing with automatic recovery from transient faults and bounded  size of needed resources. 

Our implementation also provides a number of useful building blocks, such as a reincarnation service for dealing with detectable client restarts and a self-stabilizing global reset (using an agreement protocol) that allows the preservation of the object value when the object index overflows. 


	%
\end{abstract}

\section{Introduction}

Sharing a data object among decentralized servers that provide distributed storage has been an active research topic for decades. We consider the problem of emulating a shared memory in a way that appears atomic (linearizable)~\cite{atomicity}.
Early solutions~\cite{DBLP:journals/jacm/AttiyaBD95,DBLP:conf/ac/LynchS03} do not scale well when it comes to larger data objects due to the full replication to all servers in the system. Cadambe et al.~\cite{cadambe2017coded} proposed the \emph{Coded Atomic Storage} (CAS) algorithm~\cite{cadambe2017coded}, which uses erasure coding in order to achieve data redundancy but with much lower communication cost compared with algorithms that use full replication.
Although CAS provides an efficient solution that tolerate crashes, Dolev et al.~\cite{elad} solve the same problem while considering an even more attractive notion of fault-tolerance since their solution can recover after the occurrence of \emph{transient faults}. Such faults model any violation of the assumption according to which the system was designed to operate. Their design criteria is called \emph{self-stabilization in the presence of seldom fairness} and it can tolerate both transient faults as well as more benign failures, such as crashes and packet failures, e.g., packet loss, duplication, and reordering. We will refer to their solution as CASSS (Self-Stabilizing CAS).
In~\cite{elad}, it is suggested that CASSS has similar communication costs as CAS~\cite{cadambe2017coded}. In this respect, we have implemented and empirically evaluated these solutions on PlanetLab, a planetary scale distributed infrastructure. Our results validate~\cite{elad}'s prediction. 


\subsection{Shared Memory Emulation}

The goal of emulating a shared memory is to allow the clients to access via read and write operations a shared storage in the network. By that, the service hides from the user low-level details, such as message exchange between the clients and the servers. 
%
%
As the shared data is replicated on the servers,  data consistency between the replicas (data copies) must be ensured. Atomicity (linearizability)~\cite{atomicity} is the strongest consistency guarantee and provides the illusion that operations on the distributed storage are invoked sequentially, even though they can be invoked concurrently. 
A read (resp. write) operation is invoked with a read (resp. write) request and it {\em completes} with a response (e.g., an acknowledgment). 
There are two main criteria that need to be satisfied for the atomicity property: (1)  
Any invocation of a read operation, after a write operation is completed, must return a value at least as recent as the value written by that write operation.
(2) A read operation that follows another read operation will return a value at least as recent as the value returned by the first read operation. Thus, the operations appear as sequential. 


\subsection{Fault Model}
We now discuss the fault model we consider. 
\paragraph{Benign failures.~~}
We consider message passing systems in which communication failures may occur during packet transit, such as packet loss, duplication, and reordering. However, the studied algorithms assume \emph{communication fairness}, i.e., if the sender transmits a packet infinitely often, the receiver gets this packet infinitely often.
The early solutions~\cite{DBLP:journals/jacm/AttiyaBD95,DBLP:conf/ac/LynchS03} model node failures as crashes and restrict the number $f$ of failing servers (nodes) to be less than half of the nodes in the system. We follow a similar approach but require that in the presence of transient faults, and only then, a crashed node either restarts (we call this a {\em detectable restart}) or is removed from the system via a reconfiguration service~\cite{reconfig}. Moreover, as specified in~\cite{elad}, our restriction on the number of crashes $f$ is similar to the one of CAS~\cite{cadambe2017coded}.  

\paragraph{Transient Faults.~~} 
We also consider these very rare violations of the assumptions according to which the system was designed to operate. We model their impact on the system as arbitrary changes of the state
(as long as the code stays intact). Moreover, the system starts only after the last occurrence of these very unlikely faults.      
Transient faults can, for example, be a soft error (such as a bit flip, perhaps induced by background radiation) or the very improbable event of a CRC code failing to detect a bit error in a transmitted packet. 

\paragraph{Self-Stabilization.~~}
This design criteria require recovery without external (human) intervention provides a strong fault-tolerance guarantee in that it will always recover from a transient fault.
The correctness proof of a self-stabilizing system is required to show recovery within a bounded period after the last transient fault. That is, when starting from an arbitrary system state, the system needs to exhibit legal behavior within a bounded time. 

\paragraph{Self-Stabilization in the Presence of Seldom Fairness.~~}
Dolev et al.~\cite{reconfig} proposed the following refinement of Dijkstra's design criteria of self-stabilization, which we believe to be convenient for dealing with the asynchronous nature of distributed systems.  
In the absence of transient faults, the environment is assumed to be asynchronous. Moreover, servers and clients may at any time crash. In the presence of transient faults, it is assumed that (a) all failing servers to recover eventually and (b) there is a sufficiently long period (which allows recovery) in which the system run is fair, i.e., each node makes progress infinitely often. 
%
%
%

\subsection{Related Work}


Shared memory is either single-writer and multiple-reader (SWMR), e.g., ABD~\cite{DBLP:journals/jacm/AttiyaBD95}, or multiple-writer and multiple-reader (MWMR), e.g., MW-ABD~\cite{DBLP:conf/ac/LynchS03}. 
A discussion on such non-self-stabilizing solutions is given in~\cite{DBLP:journals/eatcs/Attiya10}. 

Shared memory emulation has also been studied under dynamic server participation (e.g.,~\cite{rambo}), where changing from one server configuration to another, termed {\em reconfiguration}, requires old configuration members  to send the data to the new members; the data is replicated to all configuration members.  
 See~\cite{DBLP:journals/cacm/MusialNS14} for a survey on (non-self-stabilizing) reconfigurable solutions to memory emulation.   
ARES~\cite{Cadambe2018ARESAR} is a recent solution that supports reconfiguration of a shared memory emulation service that is based on erasure coding. The authors also present the first atomic memory service that uses erasure coding with only two rounds of message exchanges for a client operation. While combining these two create an efficient solution with liveness, even during configuration collapses, it does not consider self-stabilization.
 
Nicolaou and Georgiou~\cite{nicolaou2012} did an experimental evaluation of four non-self-stabilizing MWMR register emulation algorithms on PlanetLab. The algorithms evaluated were \textit{SWF, APRX-SWF, CwFr} and \textit{SIMPLE}. Algorithm SIMPLE is a MWMR version of ABD for quorum systems (quorums are intersecting sets of servers), similar to the one we use in this work (called MW-ABD) to compare its performance with CAS and CASSS.

Dolev~\cite{dolev2012crash} proposes a pseudo-self-stabilizing version of ABD.
%
%
There is also known ways for practically-self-stabilizing SWMR and MWMR~\cite{Alon:2011:PSA:2050613.2050617,DBLP:journals/jcss/DolevGMS18}.
Pseudo-self-stabilizing and practically-self-stabilizing systems do not guarantee bounded recovery periods, whereas self-stabilizing systems do provide such a bound. 

\subsection{Our Contribution}

We are the first to implement and evaluate via experiments a self-stabilizing algorithm for coded atomic MWMR  shared memory emulation~\cite{elad}. Our experiments show that the overhead associated with self-stabilization does not really affect the efficiency advantage associated with erasure coding.

We have also designed and implemented a reset mechanism, based on principles from~\cite{reconfig}. The reset mechanism can perform a (synchronized) global reset of the entire system while keeping the most recently written value using an agreement protocol. Additionally, we implemented a self-stabilizing reincarnation number service, which provides recyclable client identifiers, and by that helps to deal with detectable client restarts. 
	
In order to validate the analysis of~\cite{elad}, which claims comparable performance to~\cite{cadambe2017coded}, we have created pilot implementations.
Our experiments on PlanetLab show that the CAS and CASSS pilots are indeed efficient and have comparable performances with respect to operation communication latency.
The evaluation shows that our implementation of the self-stabilizing version of
CAS scales very well when increasing the number of servers and clients
respectively. More importantly,  the overhead for self-stabilization, in our experiments, is {\em constant} when compared to the implementation of the original CAS algorithm. 
The system shows almost no slowdown for data objects up to 512~KiB, and is only
slightly slower for data objects up to 1~MiB. Last but not least, the
evaluation reveals that the reset mechanism is almost as fast as a few client
operations, demonstrating CASSS's rapid recovery.
We believe that our pilots and their building blocks could be used for implementing other self-stabilizing algorithms and prototypes.


\section{System Settings and Background}
\label{ch:theory}

The system includes a network with $N$ nodes. Each node can host a client and/or a server. Servers use a gossip service for communicating among themselves. Clients interact with the shared-memory service using read and write operations. These operations include multiple communication rounds of requests and responses. Every client performs its operations sequentially, but operations can still be concurrent since clients act independently.

Servers are arranged into pairwise intersecting sets, or {\em quorums}, that together form a {\em quorum system}. The intersection property of quorums
enables information communicated to a quorum to be passed (via the common servers) to another quorum. Majorities (subsets containing a majority of the servers) form a simple quorum system (used, for example, in ABD~\cite{DBLP:journals/jacm/AttiyaBD95}). 
%
The {\em self-stabilizing} quorum system considered in this work follows the one proposed in~\cite{elad}.
We note the need for quorum systems to be self-stabilizing. This is because, for example, to the fact that client algorithms often include several phases. The clients and the servers need to be synchronized with respect to the phases, as well as the associated object version. 

Each server has access to a set that stores {\em records}; each record refers to another version of the object that a unique tag identifies. These tags also determine the causal relationship among operations, e.g., when retrieving the object's most up-to-date version. A tag has the form of $(number, clientID)$, i.e., a pair with a sequence number and the identifier of the client that is writing this version. 
More precisely, each record of algorithms CAS and CASSS is a tuple 
$(tag, data, phase)$, where  
$data$ refers to a version of an element of the coded object (or it is null) and $phase$ is the protocol `stage' that has 
written this record. 


\subsection{The MW-ABD Algorithm} 
\label{sec:abd}

The non-self-stabilizing literature includes~\cite{DBLP:journals/jacm/AttiyaBD95} for SWMR, which we refer to as \emph{ABD}, and the  MWMR counterpart~\cite{DBLP:conf/ftcs/LynchS97}, which we refer to as \emph{MW-ABD}. 
These algorithms use full replication and need to store a total of $N\times d$, where $N$ is the number of servers and $d$ is the size of one data object. Thus, a single write from a single client could occupy considerable (network and storage) resources.

MW-ABD consists of two phases. For a write, the client first carries out the \emph{query} phase and then the \emph{write} phase. During the query phase, the client sends a message to all servers, requesting their respective latest tag and associated object data. The client then waits until it has received responses from a majority of servers (or a quorum), stores the greatest tag value as $maxTag=(number, clientID)$ and ends the query phase. During the write phase, the client increments the tag counter to be $maxTagP1=(number+1, ownClientID)$, using its own identifier $ownClientID$ at the tag, and then proceeds by sending a message $(maxTagP1, data)$ to all servers. After receiving acknowledgments from a server majority, the client is assured to have successfully written a new value to the quorum system. 


The read also has a query phase from which the client retrieves both $maxTag$ and its corresponding object data. This is followed by a \emph{propagation} phase that disseminates $maxTag$ (and the object data) by sending this pair to all servers, and subsequently waiting for a majority of responses. At this point, the read operation is considered successful (returning $maxTag$'s corresponding data). 

\subsection{The CAS Algorithm}
\label{sec:cas}

\emph{Coded Atomic Storage}~(CAS)~\cite{cadambe2017coded} is based on techniques for reducing communication costs, such as erasure coding and an earlier algorithm~\cite{DBLP:conf/wdag/FanL03}, by avoiding full replication, as in ABD and MW-ABD.
CAS is a quorum based algorithm, where a quorum is any subset $Q$ of the servers, such that $|Q| = \lceil\frac{N+k}{2}\rceil$; $N$ is the number of servers and $k$ is the coding parameter deciding how many
elements are needed to decode the object value. The CAS allows for up to $f$ server failures. 

\subsubsection*{Erasure Codes}
\label{subsec:codes}

Erasure coding is a technique whereby a relatively small amount of redundant information is added to a piece of data, in order to make it robust to bit erasures. An ($N$, $k$) erasure code splits the data into $N$ coded elements which has coding applied to them such that only a subset containing $k$ elements is needed
to decode the original object value. An erasure code is said to be a \emph{maximum distance separable} (MDS) code if it has the property that the original data can be reconstructed from any $k$ of the $N$ coded elements (as
opposed to requiring one or more of the $k$ elements to be of a particular kind). The particular kind of MDS erasure coding we consider is ($N, k$)-Reed-Solomon codes, which is a group of MDS erasure codes. We note that Petig et al.~\cite{elad} show how to address privacy by merely storing on each server such data elements.


CAS builds on having ($N, k$) coding applied to the data, and distributing the $N$ coded elements to the servers in the quorum system. Since $k$ elements are required in order to decode the data, the coding
parameters have a direct effect on the quorum size. This accommodates for a flexibility in choosing between having smaller sized coded elements and better data redundancy; hence, CAS can be tweaked according to the system needs. The fraction~$r=k/N$ is called the {\em code rate}, and it measures the portion of non-redundant data in the coded elements.  



\subsubsection*{Writer's procedure}
\label{subsubsec:caswriter}
There are three phases: \emph{query}, \emph{pre-write} and~\emph{finalize}. The query is based on MW-ABD~\cite{DBLP:conf/ftcs/LynchS97}'s query, but considers only finalized records, i.e., records that their $phase$ field is `fin' (discussed next).


\textbf{Pre-write:} $p_i$'s client sends a message, $\langle (x+1, i)$, $m_j$, `pre'$\rangle$, to any server $p_j$ and waits for a quorum of replies, where $maxTag=(x, \bullet)$ is the tag retrieved from the query phase and $m_j$ holds the coded element to the server at $p_j$. 


\textbf{Finalize:} $p_i$ sends a message $\langle(x,i)$, $\bot$, `fin'$\rangle$, to all servers. After receiving a quorum of acknowledgments, the write operation is finished. The finalize phase hides the write operations that have not been seen by a quorum, since the query phase only looks at records with phase `fin'. Once the client has passed the \emph{pre-write} phase, it knows that at least a whole quorum has enough elements to reconstruct the data and therefore it can be made visible in other operations. 

\subsubsection*{Reader's procedure}
\label{sec:casreader}
There are two phases: \emph{query} and~\emph{finalize}. The first one is identical to the writer's query.


\textbf{Finalize:} client $p_i$ sends out a message $\langle (x,\bullet)$, $m_j$, `fin'$\rangle$ to all servers, where $maxTag=(x, \bullet)$ is the tag retrieved from the query phase. 
%
The client waits until a quorum has responded; each response includes a coded element corresponding to $maxTag$ (or a null if the server stored no record corresponding to $maxTag$). If at least $k$ of the responses include a coded element, the reader decodes the object value and returns it to the application. Otherwise, it just returns as an unsuccessful read. 

\subsubsection*{Server's procedures}

A server stores the different versions of the objects in records of the form ($t, w, label$), where $t$ is a tag, $w$ is a coded element and $label$ is either `pre' or `fin'. 
The server's procedures includes the event handlers corresponding to the client requests: query, pre-write and finalize (of both read and write).
Note that the algorithm clearly tolerates any writer failure (crash) whenever either no server or a quorum receives the finalize message. To the end of establishing viability of a write operation that only some servers (but not a quorum) store a finalized record, the algorithm employs a reliable gossip mechanism for disseminating among the servers tags of finalized records. This dissemination is invoked once for any arriving finalized message. 


\subsection{The CASSS Algorithm} 
\label{sec:casss}
CASSS~\cite{elad} is both self-stabilizing and privacy-preserving. We focus on CASSS's ability to recover after the occurrence of transient faults, which are violations of the assumptions according to which the system was designed to operate. This is modeled by considering transient faults that corrupt arbitrarily the system state (as long as the program code stays intact) and that they occur before the system starts to run (since they are very rare in practice). The above modeling creates the following {\em challenges}:


\begin{enumerate}[leftmargin=0.65\parindent]
\item\label{itm:corrupted} In the starting system state, the server at node $p_i$ may store tag $t_{\max}$ (in a record that its label is either `pre' or `fin'), such that due to the system asynchronous nature, it is not retrieved by any query for an arbitrary long period. The challenge is to bound the number of write operations in which stale information, such as $t_{\max}$'s record, may reside at the system without having a write that hides $t_{\max}$.

\item\label{itm:reliable} Self-stabilizing (reliable) end-to-end communications require to assume that the underlaying channels have bounded capacities~\cite[Chapter 3.2]{DBLP:books/mit/Dolev2000}. Thus, in the context of self-stabilization and asynchronous systems, the quorums that send acknowledgments to the clients  might complete write operations at a faster rate than of the \emph{reliable} gossip service delivers. It is not clear how can the writer avoid blocking in a self-stabilizing system that its communication channels are bounded (and still deliver all messages).  

\item\label{itm:bounded} All variables must be bounded, including, for example, the tag values. This means that when the system state encodes the maximum tag values, wrapping around to value zero needs not to disrupt the algorithm invariants, such as the tags' ability to order events. 
\end{enumerate}


\noindent \textbf{Addressing challenge $\mathbf{(\ref{itm:corrupted})}$.} CASSS repeatedly gossips the highest tag values that each servers has. CASSS includes in these messages the maximum tag that is part of locally stored records that their labels are `pre' and also the maximum tag of records with the labels `fin'. Also, any write operation queries for the highest `pre' tag so that the new tag of this operation is greater than all the (possibly corrupted) pre-write records in the system. (CASSS's read procedure is borrowed from CAS.) The correctness proof in~\cite{elad} demonstrates that this modification still preserves atomicity and thus CASSS addresses the first challenge.

\noindent \textbf{Addressing challenge $\mathbf{(\ref{itm:reliable})}$.}
The proof also shows that the  gossip service does not need guarantee the delivery of all messages and it is sufficient to provide eventual delivery of every message or later message, which has a higher tag value. The server then just overwrites the last received message in the buffers. 

\noindent \textbf{Addressing challenge $\mathbf{(\ref{itm:bounded})}$.}
To the end of bounding the state of each server, Dolev et al.~\cite{elad} first bounds the number of records each server stores and then bound the tag size. (Note that the client state of CAS is easy to bound and the message size is implied by the bound on its fields.)

\emph{Bounding the number of stored records} is based in the assumptions that failing clients do not restart and that each client invokes at most one instance of the write procedure. This means that at any time, a client can have at most two relevant records in any server storage (regardless of whether it is failing or not). That is, one of these records might be the one that holds the most recent object value (written by an already completed $p_i$'s operation) and the other record could be of an ongoing $p_j$'s write operation. So, any stored record older than the two most recent records from client $p_i$ is irrelevant, because it is either obsolete or stale. Thus, we can bound the number of relevant records by $2N$, where $N$ is the number of clients.
Dolev et al.~\cite{elad} reduces this bound to $N+\delta+3$ by adding to  write procedure a fourth round, labeled by `FIN', where $\delta$ is a bound on the number of read operations that occur concurrently with a write operation.


\emph{Bounding the maximum label} requires to consider the case in which the system state includes a tag that has reached its overflow value, $(\text{MAXINT},\bullet)$. Note that by choosing $\text{MAXINT}$ to be a very high value, say, $2^{64}-1$, we can grantee that such an event happens only after the occurrence of a transient fault. The solution is to let the servers to detect the presence of this overflow value and then to stop responding to queries while keeping the gossip service running . By that, the servers disseminate the overflow values in the system while abstaining from supporting new operations from installing pre-write records. This continue until the servers detect, via the gossip service, that all of them have the same maximum finalized tag value, $t_{\max}$. At that point, the algorithm in~\cite{elad} invokes a self-stabilizing global reset that allows the preservation of the object value using an agreement protocol, which our implementation bases on the one used in~\cite{reconfig}. During the reset, all clients are forced to perform also a local reset, which causes the abortion of all ongoing operations. Once the agreement procedure is terminated, the servers empty their local storages while keeping only the most recent finalized record and replacing its tag $t_{\max}$ with the initial tag value. Then, the system resumes normal operation.


\section{Implementation}
\label{ch:system}
%
%
%
%


The CASSS pilot was implemented as a library, which can be used by applications in order to provide access the read and write operations.
Calls to the functions \texttt{read()} and \texttt{write(x)} behaves just as if the service was an actual shared memory.
Calls to these functions blocks the calling process until it returns.
A successful read operation returns the data object, and a write operation blocks until it is done writing the object (and returns nothing). 





\subsection{Gossip- and Quorum-based Communications}
We used a self-stabilizing version of the token passing algorithm of~\cite[Figure~4.1]{DBLP:books/mit/Dolev2000} using UDP/IP as the basis for implementing the gossip and quorum services. 
CASSS requires the use of a self-stabilizing gossip protocol  between servers to periodically share the largest tags for each phase. Since reliably was not required, we used UDP/IP and let the arriving gossip messages to overwrite the old ones (even if the old was not delivered).
Our self-stabilizing quorum system follows the one in~\cite{elad}.
For the sake of improved performance, whenever it was required to transfer large data objects, a new TCP/IP connection was established and used. Our pilot implementation simply used a configuration file for retrieving the list of available storage servers (rather than an external directory).

\subsection{Reincarnation Service}
\label{sec:reincarnation}
CAS assumes that clients cannot resume after failing.
CASSS includes an extension that allows clients to reincarnate~\cite{elad}.
This is based on extending the client identifier to $uid$, which consists of a unique hardware address and an incarnation number.
Due to the page limit, complete details of the reincarnation service appear in Appendix~\ref{sec:reincarnation}.

The client algorithm performs a periodic task that starts with a query phase to check if its current incarnation number is up to date.
It queries all servers, and awaits responses form a quorum of servers.
The maximum value of all received incarnation numbers is calculated, and if that number differs from the current client incarnation number, a second phase is triggered.
During the second phase, the incarnation number is updated both at the client side and in the quorum system.
The client takes the maximum of the current incarnation number and all received incarnation numbers, increments that by one and sends it out to all servers.
After receiving a quorum of acknowledgments, the client knows that it has been assigned a new valid incarnation number and can thus proceed with operation as usual by updating its $uid$ accordingly.

The server algorithm has two event types that can be triggered: a query for an incarnation number and an update of an existing value. The query procedure first checks that the maximum allowed incarnation number does not appear at the server.
If there exists such a value, new incarnation number requests will be blocked in the query phase until a global reset has completed.
Otherwise, if no previous number associated with the requested hardware address exists, the default value $0$ is returned.

\subsection{Global Reset}
\label{sec:reset}

We use a global reset mechanism for restarting sequence numbers (of tags and incarnation numbers). 
This wrap around procedure is based on the ability to achieve agreement and thus we assume that all servers are alive, e.g., via a self-stabilizing service for quorum reconfiguration~\cite{reconfig}. 
We further borrow ideas from~\cite[Algorithm 3.1]{reconfig} for performing a global reset while preserving the recent object value and a mechanism for recovering from transient faults.
Due to the page limit, complete details of our global reset mechanism are given in Appendix~\ref{sec:reset}.

The servers propose their tags and then coordinate their phase transition using an agreement on the maximum tag. Once an agreement was reached, all other tags should be removed from their storage. This is referred to as the replacement phase. Since this is a self-stabilizing algorithm, it constantly checks for transient faults. If a transient fault is noticed, the algorithm cancels the replacement phase and enters a reset phase. This reset phase is used to restart the agreement process. 

	


	

\section{Evaluation Plan}
\label{ch:ev-env}

We describe the evaluation criteria and platform before the experiments. 

\subsection{Evaluation Criteria and Platform}
\label{sec:eval-criteria}

A common evaluation criteria in the field is to measure operation latency; the average time it takes for an operation to complete~\cite{nicolaou2012}. This includes both communication delay and local processing time. 
%
%
The operation latency is measured both in an isolated setting where no other clients are doing any requests and in a setting when we have different levels of base load on the servers.
For comparison, we have CAS, CASSS as well as a MW-ABD using a self stabilizing quorum system.




%
We used the PlanetLab~EU\footnote{\url{https://www.planet-lab.eu/}} platform, which provides us access to a set of virtual machines running Fedora OS version~25. A PlanetLab user gets access to a containerized instance via Linux containers (LXC). The PlanetLab~EU servers are distributed all over Europe, and since they are connected over the Internet, they do indeed provide a suitable environment to evaluate a real-world distributed application.
Because of this, an application on PlanetLab has to deal with all the real-world issues one usually runs into, such as congestion, link failures and node failures. 
In Appendix~\ref{sec:setup}, we list the PlanetLab nodes used in our experiments. 
%
%
Even though there are hundreds of machines available on the PlanetLab platform running the same operating system, they do differ in compatibility. Hence, we had to carefully pick nodes so that they had a global static IP, applications were able to bind ports and, for the case of client nodes, had the hardware support needed for the erasure coding library. 


\subsection{Experiment Scenarios}
\label{sec:test-cases}
%
%
We describe the experiment settings, and how we measure performance before giving the details of each experimental scenario. 

~~\\\subsubsection*{Baseline Settings}
\label{sec:eval-setting}
For unifying the evaluation, we often use the same baseline for each of the experiments (and otherwise note this).
%
%
The setting that all experiments proceed from is to have 15~machines in total, ten of which run one server process each and five of which run one client process each. When increasing the number of clients or servers beyond the amount of physical machines, multiple instances are put on the same physical machine. In order to guarantee a fair latency between a client and a server instance, clients processes are never placed on the same physical machine as server processes. More clients or servers than available nodes are distributed in a round-robin fashion. Operations of a client are invoked sequentially with a random delay in between.

The system is initialized by a 512~KiB data object with random data being written to the quorum system before the experiments start. Each client repeats the operation 50~times, and the fastest and slowest operations are removed in order to mitigate the effect of outliers (by pre-experiment evaluations we were able to identified 50 as a reasonable number, where experiments would complete in reasonable time, without sacrificing validity). The final operation latency result is the average of every client's average operation latency. Taking the average over all clients accounts for local variations, since different PlanetLab nodes have different conditions.
PlanetLab servers do not have any uptime guarantees, and we therefore want to allow a few servers to fail (i.e., $f > 0$). But because $k$ is bounded to be an integer value, such that $1\geq k \geq N-2f$, $f$ cannot be chosen freely. It therefore stands clear that if $f$ is constant, $N$ can never be chosen such that $k$ would be forced to be less than one. Conversely, since we want to run an experiment with as few as five servers, we have chosen $f=2$.  

~~\\\subsubsection*{Client Scalability Experiment}
\label{sec:eval-clients}
This scenario is made to evaluate how the read and write latency are affected when increasing the number of writers and readers respectively. This tests the servers' ability to handle an increase of concurrent operations. The number of failing nodes is kept constant, i.e., the quorum size is also constant. Both the reads and writes latency is measured. For reader scalability, we consider 5, 10, 15, 20, 30 and 40 readers, while having 10 writers and 10 servers. Similar numbers are used for writer scalability.  


~~\\\subsubsection*{Server Scalability Experiment}
\label{sec:eval-servers}
The server scalability experiment is constructed to evaluate in what way the read and write latencies are affected when increasing the number of servers. The number of failing nodes is kept constant, i.e., the quorum grows with the number of servers. So when the servers increase, the number of servers that a client has to access will also increase but the coded elements will be smaller. One interesting aspect to look at when increasing the number of servers is whether the effect of higher code rate trumps the effects of having a larger quorum. Both read and write latencies are measured. We use 5, 10, 15, 20 and 30 servers, while having 10 readers and 10 writers. 

~~\\\subsubsection*{Data Object Scalability Experiment}
\label{sec:eval-file}
For evaluating how the read and write latencies are affected by the object size, this experiment performs operations using increasingly large data objects. The size is increased to a maximum of 4 MiB, which was found to be enough to demonstrate the scalability. In particular, we consider objects of size 1, 32, 128, 512, 1024, 2048 and 4096 KiB.
%
%
%
The number failing node is kept constant ($f=2$), as well as the number of servers (10), which means that the quorum size is also constant. The experiment is run in isolation from other client nodes, so that scalability in increasing object sizes can be reliably measured. Both the read and write latencies are measured.\vspace{-1em} 


~~\\\subsubsection*{Reset Experiment}
\label{sec:eval-reset}
This scenario measures how long it takes for the servers to reset their local state after a transient fault. Since this part requires the participation of all servers, we do not allow any server to be unresponsive (i.e., $f=0$). Because some nodes on PlanetLab were highly unstable, it was hard to run experiments for prolonged stretches of time. Therefore, we limited the number of repetitions for the reset experiment (which was expected to take longer than the other experiments) to 20 instead of 50. For the same reason, we restricted the object size to 0.25 KiB.

Having to reset the global system state is the worst case scenario when it comes to recovery after a transient fault. The time measured is from a client pre-write phase (with a maximal tag number) until a query ends successfully. As discussed, we set $f=0$, in order to know that every server has finished the reset phase, meaning the client has to receive responses from all servers before returning.\vspace{-1em}


~~\\\subsubsection*{Overhead Experiment}
\label{sec:eval-overhead}
In this scenario, we compare the overhead of CASSS to a CAS implementations. In particular, CAS is a modified version of CASSS that does not include the fourth round (`FIN') nor does it have gossip repetitions.
In other words, this implementation uses the same number of phases and gossip messages as in~\cite{cadambe2017coded}, but, for a fair comparison, it is based on the same software components as the CASSS implementation. Here we use 10 servers (with $f=2$), one writer and one reader.




\section{Evaluation Results}
\label{ch:results}
%
We start by first looking at the two client scalability experiments, next the server scalability experiment, then the data object scalability experiment, following by
the reset time and overhead experiments. 
%
Our results show that the CASSS is efficient and scalable. Compared to CAS, it has only a constant overhead in terms of operation latency. It is efficient in storing up to 1~MiB of data, and can perform a global reset within a few seconds for systems with up to 20 servers. Given the fact that, in the absence of transient faults, CASSS performs a global reset once in every at least $2^{64}-1$ write operations, the amortized cost of this overhead is negligible.  \vspace{-1em}

~~\\\subsubsection*{Client Scalability}
\label{sec:results:number_of_clients}
\begin{figure}
	\vspace{\reduceSpace}
	\centering
	\includegraphics[scale=\figSize]{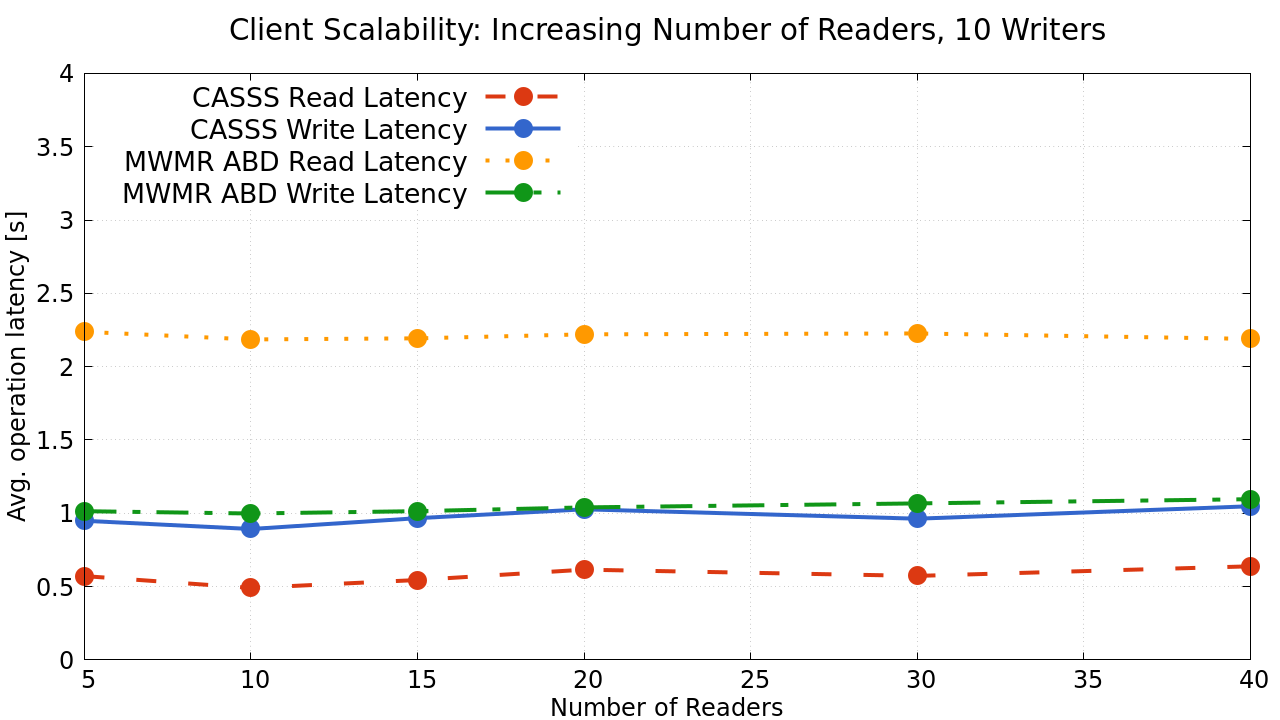}\\
	\includegraphics[scale=\figSize]{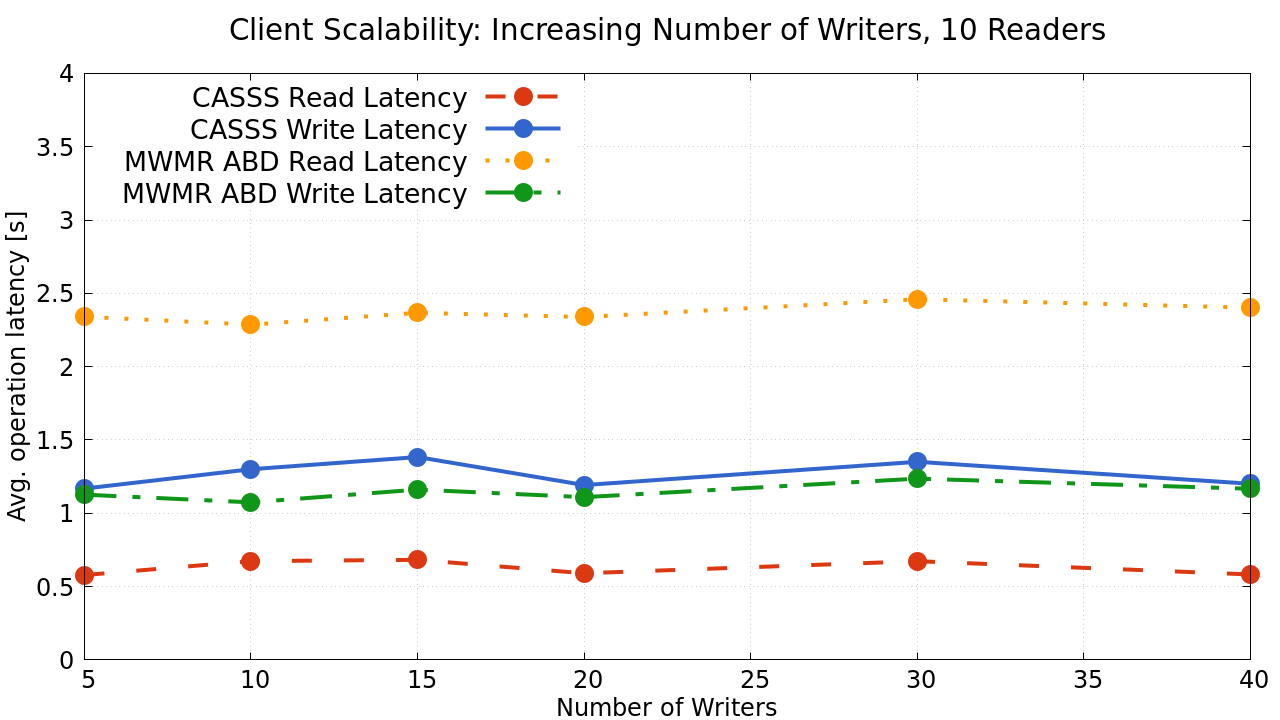}
	\caption{\label{fig:number_of_readers}\smaller{Operation latency with respect to the number of concurrent (a) readers and (b) writers.}}
	\vspace{\reduceSpace}
\end{figure}

%
Figure~\ref{fig:number_of_readers}(a) shows the result of the experiment where the number of concurrent readers was changed, and Figure~\ref{fig:number_of_readers}(b) the corresponding experiment for number of concurrent writers. Both charts show a rather flat curve, which indicates that none of the experiments reached a point where the system was overwhelmed by the number of concurrent operations.
\remove{
\begin{figure}
	\vspace{\reduceSpace}
	\centering
	\caption{\label{fig:number_of_writers}\smaller{Operation latency with respect to the number of concurrent writers.}}
	\vspace{\reduceSpace}
\end{figure}
}

%
Note the difference between operations. The fact that MW-ABD read operation is the slowest of the four is not a surprise. Not only does MW-ABD send larger messages, due to the lack of coding, but also its read operation actually transfers data twice: once to fetch the data from the servers, and once during the propagation phase. The MW-ABD and CASSS complete  write operations in about the same amount of time. While CASSS writes has two more communication rounds than MW-ABD writes, MW-ABD messages are larger due to the lack of coding. Considering the relatively short RTT between PlanetLab nodes ($\approx$ 50 msecs avg ping time), the cost of two extra rounds seems to be about as expensive as the cost of larger messages.
%
%
%
We find that CASSS reads are the fastest ones. This too was expected, since it has as few rounds as MW-ABD writes, but uses coding which decreases the message size.\vspace{-1em}

~~\\\subsubsection*{Server Scalability}
\label{sec:results:number_of_servers}
\begin{figure*}[t!] 
	\vspace{\reduceSpace}
	\centering
	\includegraphics[scale=\figSize]{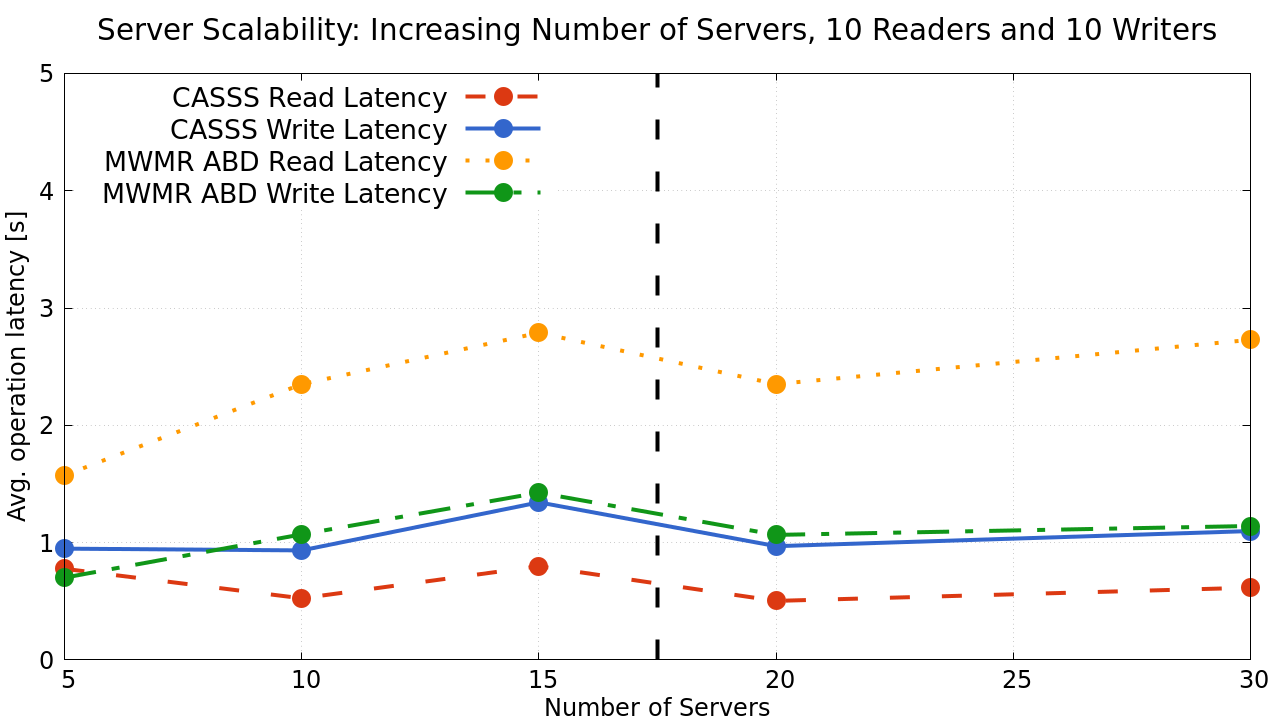}
	\caption{\label{fig:number_of_servers}\smaller{Operation latency with respect to the number of servers. The  vertical dashed line denotes the point where the parameter $f$ had to be changed.}}
	\vspace{\reduceSpace}
\end{figure*}		
Figure~\ref{fig:number_of_servers} presents the results of the servers scalability experiment. Note that with five servers, both reads and writes of CASSS and MW-ABD writes ends up at more or less the same spot. That is because with only five servers, CASSS effectively performs full replication and the CASSS quorum size is equal to majority quorum. While MW-ABD reads have fewer rounds than CASSS writes, MW-ABD reads transfer more data. This is why it the slowest of all operations.

Looking at the interval between five and ten servers, the operation latency of MW-ABD increases while the operation latency of CASSS decreases or stays the same. That is because when increasing the number of servers, the quorum size grows but so does the code rate. So while both MW-ABD and CASSS waits for responses from more servers, CASSS gains the advantage of decreased message size.
The used coding library has a limitation that $k+m \leq 32$. Thus, $f$ could not be kept at 2 for quorum systems with 20 and 30 servers. For 20 servers, $f$ had to be at least 4, and for 30 servers it had to go all the way up to 14. The point where $f$ is changed is marked by the dashed vertical line in the graph.\vspace{-1em}

\begin{figure*}[t!]
	\vspace{\reduceSpace}
	\centering
	\includegraphics[scale=\figSize]{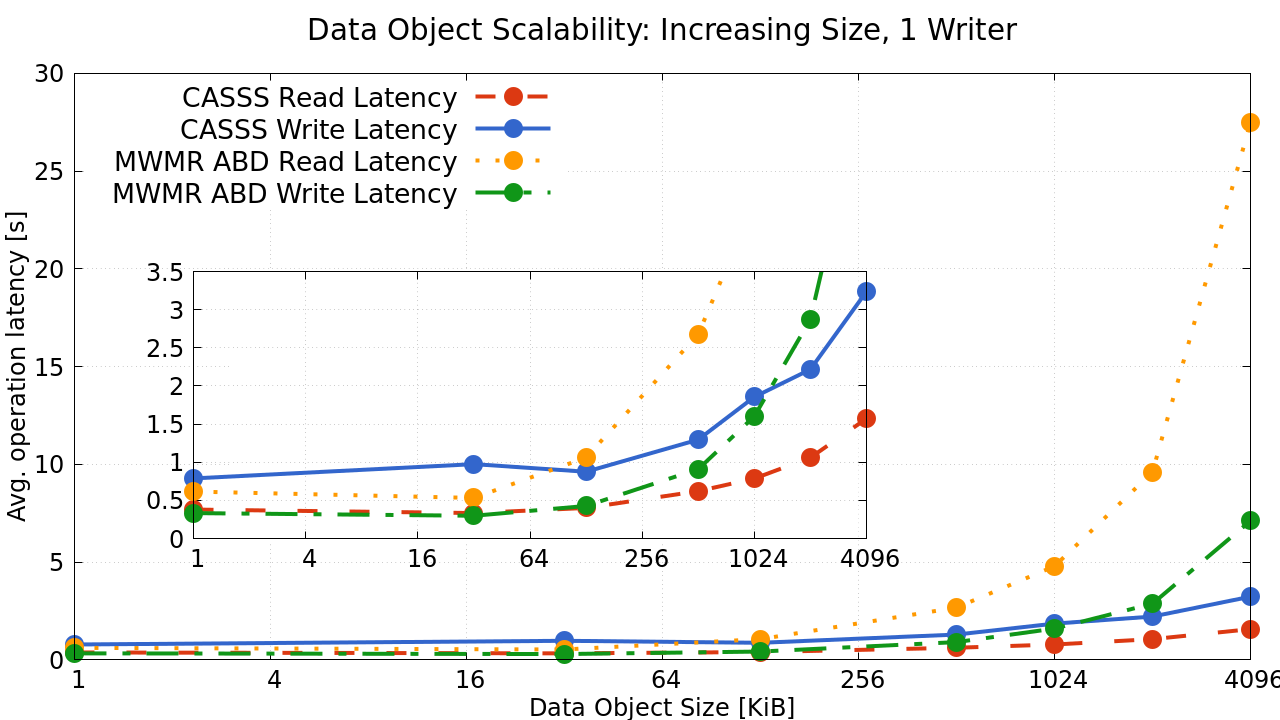}
	\caption{\label{fig:filesize}\smaller{Operation latency with respect to the size of the data object.}}
	\vspace{\reduceSpace}
\end{figure*}		

~~\\\subsubsection*{Data Object Scalability}
\label{sec:results:filesize}
Figure~\ref{fig:filesize} shows the results of the data object scalability experiment. 
(Existing solutions~\cite{DBLP:conf/wdag/FanL03} show how to transform ABD-like algorithms to more suitable implementations for large data objects.)
%
%
Up until about 1~MiB, the operation latency is fairly minimal. MW-ABD begins to escalate already at 512~KiB, but CASSS is reasonably fast all the way to 4~MiB. This is of course a consequence of the coding, which effectively reduces the message size.
%



\begin{figure*}[t!]
	\vspace{\reduceSpace}
	\centering
	\includegraphics[scale=\figSize]{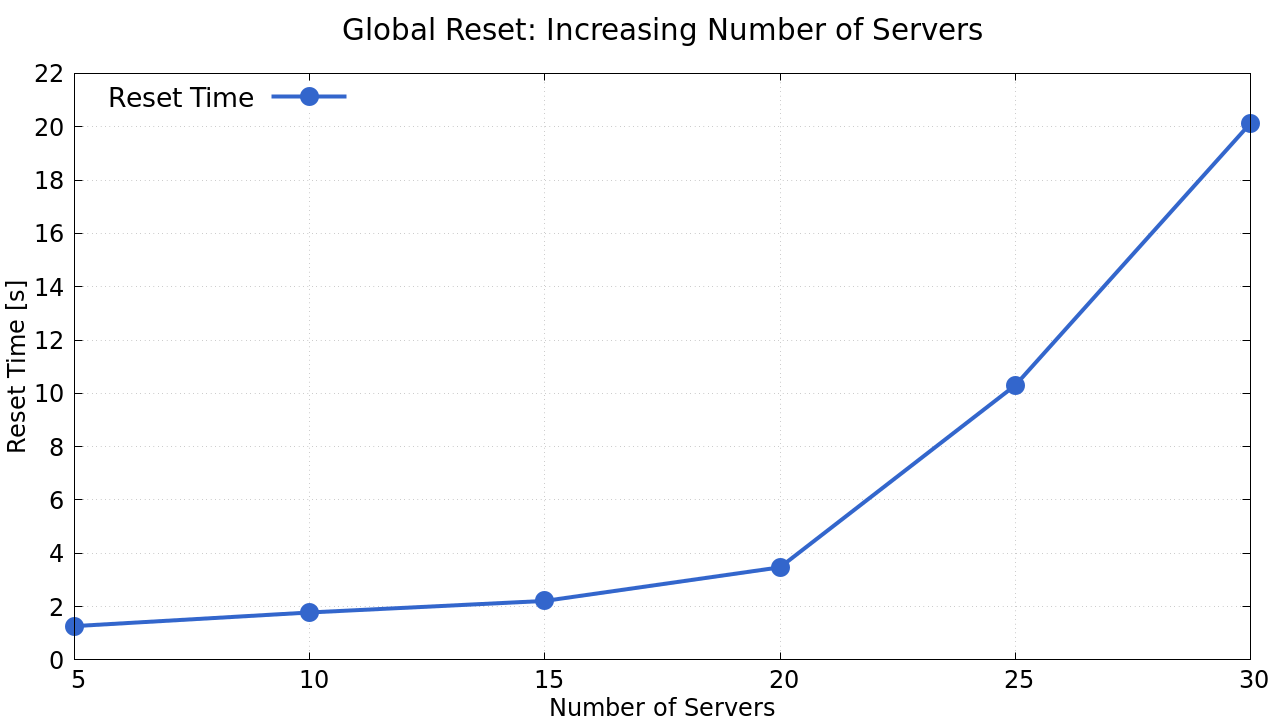}
	\caption{\label{fig:reset}\smaller{The time it takes for the Global Reset mechanism to complete, with respect to the number of servers.}}
	\vspace{\reduceSpace}
\end{figure*}		

~~\\\subsubsection*{Global Reset}
\label{sec:results:reset}
The global reset is triggered only after the occurrence of a transient fault, i.e., it is invoked very rarely. Even so, it is still important that the reset period is rather short. Figure~\ref{fig:reset} shows that, up to 20 servers, can finish reseting in the time it takes to perform two write operations, i.e., few seconds. 
%
%
As the number of servers increases, the likelihood of having to wait for slower servers increases too. If the responsiveness for a server at a given time is normally distributed, the likelihood of having one or more slow servers in the system increases exponentially.\vspace{-1em}

\begin{figure*}[t!]
	\vspace{\reduceSpace}
	\centering
	\includegraphics[scale=\figSize]{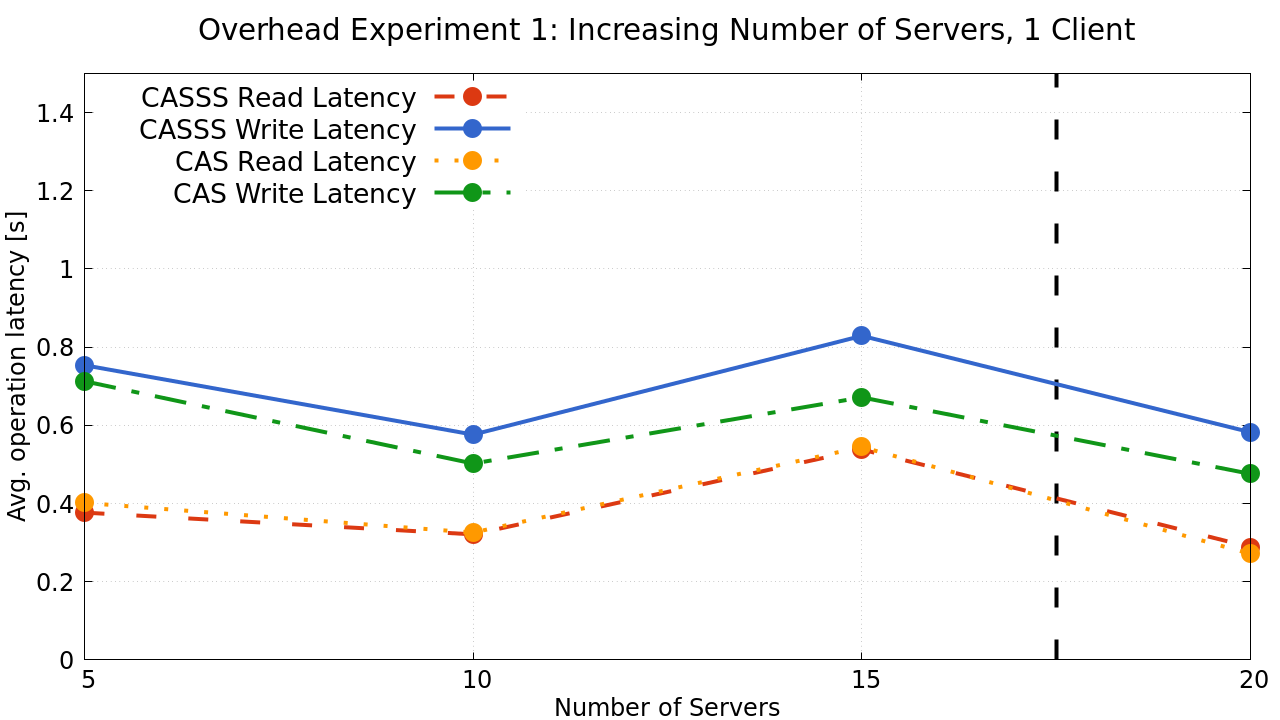}
	\caption{\label{fig:comparison_with_cas}\smaller{Comparison between the operation latency of CASSS versus the traditional CAS algorithm. The dashed vertical line denotes the point where the parameter $f$ had to be changed.}}
	\vspace{\reduceSpace}
\end{figure*}	

~~\\\subsubsection*{Overhead}
\label{sec:results:overhead}
%
Figure~\ref{fig:comparison_with_cas} reveals the overhead that the extra communication round and intensive gossiping have.
%
%
The figure has a vertical dashed line, which indicates at which point the variable $f$ was changed due to the coding library requirement discussed previously.
Note that CASSS reads and CAS reads are nearly identical. This is exactly what one would expect, since CASSS has the same number of rounds for the reads as CAS. The write operations differ slightly, and with CASSS needing one extra communication round to complete the write operation, we expected it to be slightly slower than CAS. The average ping time between the PlanetLab nodes was about 50~msecs, so the expected cost for one round is consistent with what we find in Figure~\ref{fig:comparison_with_cas}.

\section{Conclusion}
\label{sec:conclusion}
Our pilot implementation for self-stabilizing atomic MWMR coded shared memory emulation is the first to address 
benign failures, the occurrence of transient faults and the need to bound the storage size. Our experimental evaluation shows that the self-stabilization overhead is constant and the recovery period is short. We provide a self-stabilizing reset mechanism that perform a synchronized global reset of the entire system in a graceful manner that does not remove the most recently written value. Additionally, we implement a self-stabilizing reincarnation number service that provides the failing clients another chance to participate. As future work, we offer our pilot and its building blocks as the basis for other self-stabilizing systems and services, such as the ones that need quorum systems, gossip or tags.

\bibliographystyle{plain}
\bibliography{ref}

\newpage

\appendix 

\section*{\Large{Appendix}}
 
\section{Reincarnation Service}
\label{sec:reincarnation}

Reincarnation of clients is an extension to CASSS that appears in~\cite{elad}. 
Without this extension, clients cannot resume after failing, except as part of the global reset procedure.
If a client sends a request, fails and then restarts again before performing a new request, the response to the first request might be received and mistaken as the response to the new request.
That would violate correctness, and it is dealt with by the reincarnation service.

\subsection*{Extended Client Identifier}
We extend the client identifier to consists of a unique hardware address and an incarnation number.
The incarnation number is requested (and updated, if needed) at boot as well as periodically.
This way, clients do not have to delay joining until a global reset is triggered and can resume immediately.
%
%
It is important to note that a new type of sequence number, the \emph{incarnation number}, has to be introduced and it is also prone to transient faults. Such a situation is handled the same way as with a maximum tag number. If a maximum incarnation number is noticed, then new queries are blocked and a global reset is invoked.

\subsection*{Algorithm~\ref{alg:incnbr}: The Reincarnation Service}
%
%
A client has access to the $uid$ variable, which is a tuple of incarnation number and the client's globally unique hardware address. The client also has access to an interface called $qrmAccess()$, which gives access to a majority quorum of servers. A server has a first-in-first-out queue, where it stores tuples containing the hardware address and highest corresponding incarnation number for each client.
In order to bound the storage space, it is assumed that there exists an upper bound on the space of relevant hardware addresses. However, since it is a queue, the set of relevant addresses can vary over time.

The client algorithm performs a periodic task that starts with a query phase to check if its current incarnation number is up to date.
It queries all servers, and awaits responses form a quorum of servers.
The maximum value of all received incarnation numbers is calculated, and if that number differs from the current client incarnation number, a second phase is triggered.
During the second phase, the incarnation number is updated both at the client side and in the quorum system.
The client takes the maximum of the current incarnation number and all received incarnation numbers, increments that by one and sends it out to all servers.
After receiving a quorum of acknowledgments, the client knows that it has been assigned a new valid incarnation number and can thus proceed with operation as usual by updating its $uid$ accordingly.

The server algorithm has two event types that can be triggered: a query for an incarnation number and an update of an existing value. The query procedure first checks that the maximum allowed incarnation number does not appear in the server's incarnation number queue.
If there exists such a value, new incarnation number requests will be blocked in the query phase until a global reset has completed.
Otherwise, if no previous number associated with the requested hardware address exists, the default value $0$ is returned.
If the client's previous number is present, then that tuple is placed at the tail of the queue and it is sent as a response.
The update procedure is simpler and just adds the new value to the queue.
If a previous value was recorded, then the update procedure removes the old value from the queue.

\begin{algorithm*}[t!]
	\DontPrintSemicolon
	\begin{smaller}
		\KwVar{\;
			$uid$: is a tuple of hardware address and incarnation number\;
			$cntrs$: is a FIFO queue of all incarnation numbers associated with a corresponding hardware address. The size of the queue is the upper bound on the number of relevant hardware addresses allowed. New entries are included in the queue after one complete cycle.
		}
		\setcounter{AlgoLine}{0}
		\bf The client:\;
		\Upon{periodic task}{
			\bf let $incNbr \longleftarrow \max\{qrmAccess($`$cntrQry$'$)\}$\;
			\If{$incNbr \neq uid.incNbr$}{
				$newIncNbr \leftarrow \max\{i ncNbr, uid.incNbr\}+1$\;
				$qrmAccess((newIncNbr, $`$incCntr$'$))$\;
				$uid \longleftarrow \langle hwAddr, newIncNbr \rangle $\;
			}
		}
		\bf The server:\;
		\Upon{cntrQry arrival {\bf from} $p_{j}$'s client {\bf to} $p_{i}$'s server}{
			\lIf{$maxIncNbr \in cntrs$}{\bf return}
			\If{$\langle j, \bullet \rangle \in cntrs$}
			{$cntrs.add(cntrs.remove(j))$\;
				$reply(j,(cntrs.get(j).incNbr, $`$cntrQry$'$))$}
			\lElse{$reply(j,(0, $`$cntrQry$'$))$}
			
		}
		\Upon{(newIncNbr,`incCntr') arrival {\bf from} $p_{j}$'s client {\bf to} $p_{i}$'s server}{
			\lIf{$\langle j, \bullet \rangle \in cntrs$}
			{$cntrs.remove(j)$}
			$cntrs.add( \langle hwAddr_j, newIncNbr \rangle)$
		}
		\caption{\label{alg:incnbr}\smaller{Algorithm for reincarnation service. Code for $p_i$'s client/server.}}
	\end{smaller}
\end{algorithm*}

\section{Global Reset}
\label{sec:reset}

A global reset mechanism is needed to reset sequence numbers (of tags and incarnation numbers) and wrap around to a default value. 
%
%
This wrap around procedure is based on the ability to achieve agreement and thus we assume that all servers are alive, e.g., via a self-stabilizing service for quorum reconfiguration~\cite{reconfig}. 
%
%
%
Algorithm~\ref{alg:global-reset} borrows its core ideas from~\cite[Algorithm 3.1]{reconfig} for performing a global reset while preserving the recent object value and a mechanism for recovering from transient faults.

\begin{algorithm*}[t!]
	\DontPrintSemicolon
	\begin{smaller}
		\KwVar{\;
			$prp[]$,
			$all[]$,
			$echoAnswers[]$,
			$allSeenProcessors$,
			$dftlPrp = \langle 0, \bot\rangle$
		}
		\KwFunc{$propose(tag)=
			\{$\lIf{$enableReset()$}{
				$(prp[i], all[i]) \leftarrow (\langle 1,tag \rangle, false)\}$}
		}
		\KwMacro{$enableReset()=
			\KwRet
			(\nexists p_k \in config : (prp[k] = \bot) \newline \vee ((prp[k], all[k]) \neq (dfltPrp, true)))$
		}
		\KwMacro{$prpSet(val)=$
			\lForEach{$p_{k} \in config$}{$(prp[k], all[k]) \leftarrow (val, false)$}
		}
		
		\KwMacro{$modMax()=$ \leIf{$Phs = \{0, 1\}$}
			{\KwRet $\max Phs$}
			{\KwRet $prp[i].phase$, \bf where $Phs=\{ prp[k].phase\}_{p_k \in config}$}
		}
		
		\KwMacro{$degree(k)=$ \KwRet ($2 \cdot prp[k].phase + | \{1 : myAll(k)\}|$)
			
		}
		
		\KwMacro{$corrDeg(k, k')=$ \KwRet ($\{\{degree(k), degree(k')\} \in \{\{x,x\}, \{x, x+1 \bmod 6\}, \{x,
			x+2 \bmod 6\} : x \in \{0,\dots,5\}\}$)
		}
		
		\KwMacro{$maxPrp()=$
			\leIf{ $\{(degree(k) - degree(i)) \bmod 6\}_{p_k \in config} \nsubseteq \{0, 1\}$ }
			{
				\KwRet $prp[i]$
			}
			{
				\KwRet $\langle modMax(), max_{lex}\{prp[k].tag\}_{p_k \in config}\rangle$
			}
		}
		\KwMacro{$myAll(k)=$
			\KwRet $(all[k] \vee (\exists p_{l} \in allSeenProcessors : prp[i].phase + 1 \bmod
			3 = prp[l].phase))$
		}
		\KwMacro{$greaterOrEqual(k)=$
			\KwRet $(prp[i].phase +1) \bmod 3 = prp[k].phase \vee prp[i] = prp[k]$
		}
		\KwMacro{$echoNoAll(k)=$
			\KwRet $(prp[i]=echoAnswers[k].prp) \wedge greaterOrEqual(k)$
		}
		\KwMacro{$echo(k)=$
			\KwRet $(\{(prp[i],all[i])\}=\{echoAnswers[k]\}) \wedge greaterOrEqual(k)$
		}
		
		\KwMacro{$increment(prp)=$
			\bf case ($prp.phase$) of $1$: return $(\langle2, prp.tag\rangle, false)$; 2: return $(dfltPrp, false)$; else return $(prp[i], all[i])$;
		}
		
		\KwMacro{$allSeen()=
			(all[i] \wedge config \subseteq (allSeenProcessors \cup \{p_{i}\}))$
		}
		\KwMacro{$proposalSet=
			\{prp[k].tag: \exists p_{k'} \in config: prp[k'] = \langle 2, \bullet \rangle\}_{p_k \in config}$
		}
		\Doforever{}{\label{ln:reset-loop}
			\lForEach{$p_{k} \in config : all[k]$}{$allSeenProcessors
				\leftarrow allSeenProcessors \cup \{p_{k}\}$}
			\If{
				$
				(\exists p_k : ((prp[k] = \langle 0, s\rangle) \wedge (s \neq \bot))
				\vee
				(\exists p_k, p_{k'} \in config : \neg corrDeg(k, k'))
				\vee
				(\{p_k \in config : prp[i].phase + 1 \bmod 3 = prp[k].phase\} \nsubseteq allSeenProcessors)
				\vee
				(|proposalSet| > 1)
				\vee
				((\exists p_{k} \in config : prp[k] = \bot) \wedge (prp[i] \neq \{dfltPrp\})))
				$
				\label{ln:reset-iftransient}
			}
			{ $prpSet(\bot)$ \label{ln:reset-setbot}}
			\label{ln:reset-updateallk}
			\If{
				$(prp[i] = \bot \wedge all[i])$
			}{
				$prp[i] \leftarrow dfltPrp$
				\label{ln:reset-setdflt}
			}
			$(prp[i], all[i]) \leftarrow (maxPrp(), \bigwedge_{p_{k} \in config}(echoNoAll(k)))$\;
			\If{$(Prps \neq \{dfltPrp\} \wedge \nexists x\in Prps : x=\bot)$,
				where $Prps=\{prp[k]\}_{p_k \in config}$
				\label{ln:reset-ifnoreset}
			}{
				\lIf{$allSeen() \wedge \bigwedge_{p_{k} \in config}(echo(k))$}{$((prp[i], all[i]), allSeenProcessors) \leftarrow (increment(prp[i]),\emptyset)$}
				\lIf{$prp[i].phase = 2$}{$localReset(prp[i].tag)$}
				\label{ln:reset-localreset}
			}
		}
		\caption{\label{alg:global-reset}\smaller{Algorithm to perform a global reset, using coordinated phase transitions. Code for the server at node $p_i$.}}
	\end{smaller}
\end{algorithm*}

\subsection*{Global Reset Algorithm}

%
%
The servers propose their tags and then coordinate their phase transition via an agreement on the maximum tag. Once an agreement was reached, all other tags should be removed from their storage (using the $localReset()$ procedure). This is referred to as the replacement phase. Since this is a self-stabilizing algorithm, it constantly checks for transient faults. If a transient fault is noticed, the algorithm cancels the replacement phase and enters a reset phase. This reset phase is used to restart the replacement process.
%
%

Algorithm~\ref{alg:global-reset} use the variables: $prp$, $all$, $echoAnswers$, $allSeenProcessors$, $config$ and $dfltPrp$. The lists $prp$ and $all$ store received proposals and whether or not all have seen their proposals respectively.
The list $echoAnswers$ holds the latest value that a processor has sent, which has also been acknowledged by the servers.
The set $allSeenProcessors$ are used to gather all servers that have reported that everyone has seen their proposal.
Addresses to all participants are in $config$ and $dfltNtf$ is a default proposal used when there is no wrap around currently in progress.
The mechanism's correctness proof appears in~\cite{reconfig}.

\subsection*{Recovering from Transient Faults}


A reset of the proposals is done upon a transient fault is detected (line~\ref{ln:reset-iftransient}).
This process is triggered by line~\ref{ln:reset-setbot} and results in a $\bot$ in every $prp_i[k] : p_k \in config$. The goal of this is to stop any ongoing global reset procedure and start over from a state where every processor $p_i$ has $prp_i[i] = dfltPrp$. During the reset phase no processor can propose a new record with a call to $propose(tag)$ due to being blocked by the macro $enableReset()$ until the reset phase has finished.


\section{PlanetLab Setup}
\label{sec:setup}
Table~\ref{tab:planetlab-servers} lists the PlanetLab nodes, which were used as servers, and Table~\ref{tab:planetlab-clients} lists the PlanetLab nodes, which were used as clients.
As already mentioned, although there are hundreds of machines available on the Planet Lab platform and they run the same operating system, they do differ in compatibility. Hence, we had to carefully pick nodes so that they had a global static IP, that applications were able to bind to ports and, for the case of client nodes, had the hardware support needed for the erasure coding library.

\begin{table}[h!]
	\centering
	\begin{tabu}{lll}
		\rowfont\bf
		Hostname                              & TLD & IP Address\\
		cse-yellow.cse.chalmers.se            & se  & 129.16.20.70\\
		planetlab-1.ing.unimo.it              & it  & 155.185.54.249\\
		planetlab-2.cs.ucy.ac.cy              & cy  & 194.42.17.164\\
		planetlab-2.ing.unimo.it              & it  & 155.185.54.250\\
		planetlab2.upm.ro                     & ro  & 193.226.19.31\\
		ple1.cesnet.cz                        & cz  & 195.113.161.13\\
		ple1.planet-lab.eu                    & eu  & 132.227.123.11\\
		ple2.planet-lab.eu                    & eu  & 132.227.123.12\\
		ple4.planet-lab.eu                    & eu  & 132.227.123.14\\
		ple44.planet-lab.eu                   & eu  & 132.227.123.44\vspace{.5em}
	\end{tabu}
	\caption{The ten PlanetLab nodes which were used for servers in the
		experiments.}
	\label{tab:planetlab-servers}
\end{table}

\begin{table}[h!]
	\centering
	\begin{tabu}{lll}
		\rowfont\bf
		Hostname                  & TLD & IP Address\\
		pl1.uni-rostock.de        & de  & 139.30.241.191\\
		pl2.uni-rostock.de        & de  & 139.30.241.192\\
		planet4.cs.huji.ac.il     & il  & 132.65.240.103\\
		planetlab11.net.in.tum.de & de  & 138.246.253.11\\
		planetlab13.net.in.tum.de & de  & 138.246.253.13\vspace{.5em}
	\end{tabu}
	\caption{The five PlanetLab nodes which were used for clients in the experiments.}
	\label{tab:planetlab-clients}
\end{table}

\end{document}